\documentclass[twocolumn,preprintnumbers]{revtex4}
\usepackage{amsmath}

\def\no{\nonumber}
\def\LQCD{\Lambda_\text{QCD}}
\def\Br{\text{Br}}
\def\gev{\text{GeV}}
\def\mev{\text{MeV}}

\begin{document}

\preprint{TTP09-22, SFB/CPP-09-62}

\title{\boldmath $B^-\to\pi^-\pi^0/\rho^-\rho^0$ to NNLO in QCD
  factorization\unboldmath} 

\author{Guido~Bell${}^1$ and Volker~Pilipp${}^2$}

\affiliation{
$^1$Institut f\"ur Theoretische Teilchenphysik, 
Universit\"at Karlsruhe,
76128 Karlsruhe, Germany\\
$^2$
Albert Einstein Center for Fundamental Physics,
Institute for Theoretical Physics, 
University of Bern, 
Sidlerstrasse 5, 
3012 Bern, Switzerland
}

\date{\today}

\begin{abstract}\noindent
The approximate tree decays $B^-\to\pi^-\pi^0/\rho^-\rho^0$ may serve as
benchmark channels for testing the various theoretical descriptions of
the strong interaction dynamics in hadronic $B$ meson decays. The ratios
of hadronic and differential semileptonic $B\to\pi\ell\nu/\rho\ell\nu$
decay rates at maximum recoil provide particularly clean probes of the
QCD dynamics. We confront the recent NNLO calculation in the QCD
factorization framework with experimental data and find support for the
factorization assumption. A detailed analysis of all tree-dominated
$B\to\pi\pi/\pi\rho/\rho\rho$ decay modes seems to favour somewhat
enhanced colour-suppressed amplitudes, which may be accommodated in QCD
factorization by a small value of the first inverse moment of
the $B$ meson light-cone distribution amplitude,
$\lambda_B\simeq250~\mev$. Precise measurements of the $B\to\rho\ell\nu$
spectrum could help to clarify this point. 
\end{abstract}

\maketitle

\section{Introduction}

A wealth of observables at current and future $B$ physics experiments is
related to exclusive hadronic decay modes. $B$ decays into a pair of
light (charmless) mesons are of particular phenomenological interest as
they are mediated by rare flavour-changing $b\to q$~$(q=u,d,s)$ quark
transitions and the interference of several weak decay amplitudes may
induce sizeable CP-violating effects. 

The complicated strong interaction dynamics in ha\-dronic decays poses a
serious challenge for accurate theoret\-ical predictions. In recent
years systematic methods have been developed, which are based on the
factorization of short- and long-distance effects in the heavy quark
limit $m_b\gg\LQCD$. The theoretical concepts are known as QCD
factorization (QCDF)~\cite{QCDF}, soft-collinear effective theory
(SCET)~\cite{SCET} and the pQCD approach~\cite{pQCD}.  

In this letter we consider the decays $B^-\to\pi^-\pi^0/\rho^-\rho^0$
within the QCDF framework, which is based on the statement that the
hadronic matrix elements of the operators in the effective weak
Hamiltonian simplify in the heavy quark limit according to~\cite{QCDF} 
\begin{align}
\label{eq:fact}
& \langle M_1 M_2 | Q_i | \bar{B} \rangle
\simeq \;
F^{B M_1}(0) f_{M_2} \!\!
\int du \; T_{i}^I(u) \phi_{M_2}(u)
\\
&
+ \hat{f}_{B} f_{M_1} f_{M_2} \!\!
\int d\omega dv du \; T_{i}^{II}(\omega,v,u)
\phi_B(\omega) \phi_{M_1}(v) \phi_{M_2}(u).\no
\end{align}
The factorization formula implies, on the one hand, that the theoretical
prediction requires non-trivial hadronic input parameters, such as decay
constants $f$, moments of light-cone distribution amplitudes $\phi$ and
form factors $F$, which encode all long-distance effects in the limit
$m_b\to\infty$. The power of the decomposition in (\ref{eq:fact}) lies,
on the other hand, in the fact that it provides the path to a systematic
implementation of radiative corrections. The short-distance
hard-scattering kernels $T_i^{I,II}$ are perturbatively calculable and
currently being worked out to next-to-next-to-leading order
(NNLO)~\cite{NNLO:T2:tree,NNLO:T2:peng,NNLO:T1:tree}. 

The NNLO calculation is to date incomplete, but a subset of
hard-scattering kernels, which specify the so-called topological tree
amplitudes, has recently been determined to
NNLO~\cite{NNLO:T2:tree,NNLO:T1:tree}. This allows us to present the
first complete NNLO prediction within the QCDF framework for the decays
$B^-\to\pi^-\pi^0/\rho^-\rho^0$, which are pure tree decays in the
excellent approximation that small electroweak penguin amplitudes are
neglected~\footnote{In our numerical analysis the electroweak penguin
  amplitudes will be included in the NLO approximation.}.  

As the considered decays are likely to be dominated by their standard
model contribution, they may serve as benchmark channels for testing the
various theoretical descriptions of the strong interaction dynamics in
hadronic $B$ decays. By normalizing the hadronic decay rates to their
semi\-leptonic counterparts $B\to\pi\ell\nu/\rho\ell\nu$ at maximum
recoil, most of the theoretical uncertainties from hadronic input
parameters and $|V_{ub}|$ drop out and one obtains precision observables
for testing the QCD dynamics of the topological tree amplitudes. We
confront the NNLO prediction in QCDF with experimental data and find
support for the factorization assumption. We also take a look at the
other tree-dominated $B\to\pi\pi/\pi\rho/\rho\rho$ decay modes and
conclude that the colour-suppressed tree amplitudes seem in general to
be somewhat enhanced, which may hint at a smaller value of the first
inverse moment of the $B$ meson light-cone distribution amplitude,
$\lambda_B\simeq250~\mev$. We conclude our analysis with a comment on
the so-called $B\to\pi\pi$ puzzle.

\section{Tree amplitudes}

The decay amplitudes for hadronic $B$ meson decays are conveniently
parameterized by a set of topological amplitudes, which contain
short-distance QCD and some electro\-weak effects. In the notation
of~\cite{Beneke:2003zv} they read  
\begin{align}
& {\cal{A}}(B^-\to\pi^-\pi^0) =
\Big[ \lambda_u  \Big( \alpha_1 + \alpha_2
+ \frac32 \alpha^u_{3,\text{EW}} + \frac32 \alpha^u_{4,\text{EW}} \Big)
\no\\
&\qquad
+ \lambda_c  \Big(
\frac32 \alpha^c_{3,\text{EW}} + \frac32 \alpha^c_{4,\text{EW}} \Big)
\Big] \frac{A_{\pi\pi}}{\sqrt{2}}
\label{eq:amplitude}
\end{align}
with $\lambda_p=V_{pb}V_{pd}^*$ and $A_{\pi\pi}=i G_F/\sqrt{2} \,m_B^2
f_\pi F_+^{B\pi}(0)$ and similarly for $B^-\to\rho^-\rho^0$ with
$f_\pi\to f_\rho$, $F_+^{B\pi}\to A_0^{B\rho}$ and $\alpha_i(\pi\pi) \to
\alpha_i(\rho\rho)$. Whereas the electro\-weak penguin amplitudes
$\alpha^p_{3/4,\text{EW}}$ are currently known for
$B\to\pi\pi/\pi\rho/\rho\rho$ to NLO~\cite{QCDF,Beneke:2003zv,QCDF:VV},
the tree amplitudes $\alpha_{1,2}$ have recently been determined for
$B\to\pi\pi$ to NNLO~\cite{NNLO:T2:tree,NNLO:T1:tree}. From the
projection properties of the leading-twist $\pi$ and $\rho$ wave
functions, we find that the respective expressions for
$B\to\pi\rho/\rho_L\rho_L$ are identical ($L$ refers to the longitudinal
polarization). We in particular do not consider decays into transverse
$\rho$ mesons, which cannot be described model-independently as they do
not factorize.    

We evaluate the tree amplitudes with 3-loop running coupling constant
and NNLL Wilson coefficients~\cite{Wilson:NNLL} of the operators in the
weak effective Hamiltonian (we use the operator basis
from~\cite{CMM}). The spectator scattering part ($T_i^{II}$) receives
contributions from two perturbative scales, $\mu_h\sim m_b$ and
$\mu_{hc}\sim (\LQCD m_b)^{1/2}$, which gives rise to logarithms $\ln
m_b/\LQCD$ that we resum via renormalization group equations in SCET to
LL approximation. Other scale dependent quantities are treated as
described in~\cite{NNLO:T1:tree}, except for the parameters of the $B$
meson wave function, which we evolve with fixed order relations as their
evolution from their input scale does not induce parametrically large
logarithms.  

We also include certain power corrections to the tree amplitudes that
are related to subleading-twist wave functions of the light mesons. As
these chirally enhanced contributions do not factorize, we use the model
proposed in~\cite{QCDF} to estimate their size.

\begin{table}[t!]
\begin{tabular}{|cc|cc|} \hline
\hspace*{2cm}&\hspace*{2cm}&\hspace*{2cm}&\hspace*{2cm} \\[-1.4em]
Parameter & Value & Parameter & Value\\
\hline
$f_\pi$ &
$0.131$ &
$\Lambda^{(n_f=5)}_{\overline{\text{MS}}}$ &
$0.204$
\\
$f_\rho$ &
$0.216\pm0.005$ &
$\Lambda^{(n_f=4)}_{\overline{\text{MS}}}$ &
$0.283$
\\
$f_B$ &
$0.200\pm0.020$ &
$m_{b,\text{pole}}$ &
$4.8$
\\
$F_+^{B\pi}(0)$ &
$0.26\pm0.04$ &
$m_{c,\text{pole}}$ &
$1.4\pm0.2$
\\
$A_0^{B\rho}(0)$ &
$0.30\pm0.05$ &
$|V_{cd}|$ &
$0.230\pm0.011$
\\
$a_2^\pi$ &
$0.25\pm0.15$ &
$10^3|V_{cb}|$ &
$41.2\pm1.1$
\\
$a_2^\rho$ &
$0.15\pm0.15$ &
$10^3|V_{ub}|$ &
$3.95\pm0.35$
\\
$\lambda_B$ &
$0.400\pm0.150$ &
$\gamma$ &
$(70\pm20)^{\circ}$
\\
$\sigma_1$ &
$1.5\pm1.0$ &
$\mu_h$ &
$4.8^{+4.8}_{-2.4}$
\\
$\sigma_2$ &
$3\pm2$ &
$\mu_{hc}$ &
$1.5^{+1.5}_{-0.7}$
\\
\hline
\end{tabular}
\caption{\label{tab:input}
List of input parameters (in units of GeV or dimensionless). Scale
dependent quantities refer to $\mu=1~\gev$.} 
\end{table}

Our theoretical input parameters are listed in Table~\ref{tab:input}. We
deduced our default values for the hadronic parameters from recent
lattice and sum rule calculations (where available). In general the
parameters related to the pion ($f_\pi$,
$a_2^\pi$\cite{a2:pi,Boyle:2008nj}, $F_+^{B\pi}$\cite{Fplus}) are better
determined than the ones related to the rho meson ($f_\rho$\cite{frho},
$a_2^\rho$\cite{Ball:2006nr,Boyle:2008nj},$A_0^{B\rho}$\cite{A0}). While
there exists a large number of calculations for the $B$ meson decay
constant $f_B$\cite{fB}, less is known about the moments of the $B$
meson wave function ($\lambda_B,\sigma_{1,2})$\cite{lambdaB}. Our value
for $\lambda_B$ is based on a QCD sum rule calculation and on estimates
from the operator product expansion, accounting for recent claims that
higher dimensional operators lower the value of $\lambda_B$ (last paper
of~\cite{lambdaB}). 

We estimate the size of higher order perturbative corrections by varying
the factorization scales $\mu_h$ and $\mu_{hc}$ independently within the
ranges specified in Table~\ref{tab:input}. On the other hand we evaluate
the non-factorizable power corrections at a fixed scale
$\mu_0=1.5~\gev$. The latter introduce certain model parameters
($\rho_H,\phi_H$) and some additional hadronic parameters. We use
$(\bar{m}_u+\bar{m}_d)(2\gev)=8~\mev$, $\bar{m}_b(\bar{m}_b)=4.2~\gev$
and $f_\rho^\perp(1\gev)=165~\mev$.

This brings us to our NNLO prediction of the colour-allowed ($\alpha_1$)
and colour-suppressed ($\alpha_2$) tree amplitudes. In the
$B\to\pi\pi/\rho_L\rho_L$ channels we obtain 
\begin{widetext}
\begin{align}
% uncertainties refer to hadronic/scale/power
%
\alpha_1(\pi\pi) &  \;=\;
1.013\,^{+0.017}_{-0.031}\,^{+0.008}_{-0.011}\,^{+0.014}_{-0.014}
+\Big(+
0.027\,^{+0.006}_{-0.010}\,^{+0.020}_{-0.013}\,^{+0.014}_{-0.014} \Big)
i
\;=\;
1.013^{+0.023}_{-0.036}
+ \Big(+ 0.027^{+0.025}_{-0.022}  \Big) i,
\no\\
\alpha_2(\pi\pi) &  \;=\;
0.195\,^{+0.119}_{-0.066}\,^{+0.025}_{-0.025}\,^{+0.055}_{-0.055}
+\Big(-
0.101\,^{+0.017}_{-0.010}\,^{+0.021}_{-0.029}\,^{+0.055}_{-0.055} \Big)
i
\;=\;
0.195^{+0.134}_{-0.089}
+ \Big(- 0.101^{+0.061}_{-0.063}  \Big) i,
\no\\
\alpha_1(\rho_L\rho_L) &  \;=\;
1.017\,^{+0.017}_{-0.029}\,^{+0.010}_{-0.011}\,^{+0.014}_{-0.014}
+\Big(+
0.025\,^{+0.007}_{-0.013}\,^{+0.019}_{-0.013}\,^{+0.014}_{-0.014} \Big)
i
\;=\;
1.017^{+0.024}_{-0.034}
+ \Big(+ 0.025^{+0.025}_{-0.023}  \Big) i,
\no\\
\alpha_2(\rho_L\rho_L) &  \;=\;
0.177\,^{+0.110}_{-0.063}\,^{+0.025}_{-0.029}\,^{+0.055}_{-0.055}
+\Big(-
0.097\,^{+0.021}_{-0.012}\,^{+0.021}_{-0.029}\,^{+0.055}_{-0.055} \Big)
i
\;=\;
0.177^{+0.126}_{-0.089}
+ \Big(- 0.097^{+0.062}_{-0.063}  \Big) i,
\end{align}
\end{widetext}
where the uncertainties in the intermediate results stem from the
variation of hadronic input parameters, higher order perturbative
corrections and the considered model for power corrections,
respectively, which have been added in quadrature for our final error
estimate. 

We see, on the one hand, that the colour-allowed tree amplitudes
$\alpha_1$ can be computed precisely in the factorization framework. The
colour-suppressed amplitudes $\alpha_2$ suffer, on the other hand, from
substantial theoretical uncertainties. The problem is related to certain
cancellations between various perturbative contributions, which make the
real parts particularly sensitive to the spectator scattering mechanism
which is proportional to the hadronic ratio
$f_{M_1}\hat{f}_B/\lambda_BF^{B M_1}(0)$. Our poor knowledge of the $B$
meson parameter $\lambda_B$ in particular translates into the
uncertainties $\,^{+0.107}_{-0.049}$ and $\,^{+0.096}_{-0.043}$ for the
real parts of $\alpha_2(\pi\pi)$ and $\alpha_2(\rho_L\rho_L)$,
respectively.

\section{Branching ratios}

The branching ratios of $B^-\to\pi^-\pi^0/\rho^-\rho^0$ depend in
addition on electro\-weak penguin amplitudes,
cf.~(\ref{eq:amplitude}). These amplitudes have not yet been determined
to NNLO~\footnote{Partial NNLO results of the penguin amplitudes from
  spectator scattering can be found in~\cite{NNLO:T2:peng}.}, but their
numerical values are rather small
($|\alpha^p_{3/4,\text{EW}}|\lesssim0.01$). As they are not CKM-enhanced
in tree-dominated decays, it is consistent to treat these amplitudes in
the NLO approximation. The explicit NLO results can be found
in~\cite{QCDF,Beneke:2003zv,QCDF:VV} (they are formulated in a different
operator basis of the effective Hamiltonian). The CP-averaged branching
ratios become  
\begin{align}
% uncertainties refer to CKM/hadronic/scale/power
%
10^6\Br(B^-\to\pi^-\pi^0) &  \;=\;
6.22\,^{+1.14}_{-1.05}\,^{+2.03}_{-1.65}\,^{+0.16}_{-0.18}\,^{+0.43}_{-0.42}
\no\\
& \;=\;
6.22^{+2.37}_{-2.01},
\no\\
10^6\Br(B^-\to\rho_L^-\rho_L^0) &  \;=\;
21.0\,^{+3.9}_{-3.5}\,^{+7.4}_{-6.1}\,^{+0.5}_{-0.7}\,^{+1.5}_{-1.4}
\no\\
& \;=\;
21.0^{+8.5}_{-7.3},
\end{align}
where the uncertainties in the intermediate results are due to CKM
parameters, hadronic parameters, higher order perturbative corrections
and non-factorizable power corrections, respectively. 

Our NNLO results are in good agreement with experimental
data~\cite{HFAG,rhoLrhoL}, 
\begin{align}
10^6\Br(B^-\to\pi^-\pi^0) |_\text{exp} &  \;=\;
5.59^{+0.41}_{-0.40},
\no\\
10^6\Br(B^-\to\rho_L^-\rho_L^0) |_\text{exp} &  \;=\;
22.5^{+1.9}_{-1.9},
\end{align}
i.e.~the experimental values are reasonably well reproduced by the
\emph{central values} of our NNLO prediction, which is based on the
input parameters from Table~\ref{tab:input}. One should keep in mind,
however, that we could also have obtained similar numbers for the
branching ratios with rather different values of the tree amplitudes,
the form factors and $|V_{ub}|$. As we discuss in the following section,
a much stronger test of the factorization assumption can be obtained by
considering the ratios of hadronic and differential semileptonic decay
rates, where the dependence on the form factors and $|V_{ub}|$ drops out
to a large extent. 

We may also take a look at the other tree-dominated
$B\to\pi\pi/\pi\rho/\rho\rho$ decay modes. We emphasize that the NNLO
calculation of these branching ratios is to date still incomplete, since
the QCD penguin amplitudes have not yet been determined to NNLO (this is
why we do not discuss CP asymmetries in this letter). These modes also
differ conceptually from $B^-\to\pi^-\pi^0/\rho^-\rho^0$ in the sense
that they receive contributions from weak annihilation, which
constitutes another class of non-factorizable power corrections. We
again use the model from~\cite{QCDF} to estimate their size.

Our results for the CP-averaged branching ratios are shown in
Table~\ref{tab:BRs}. Apart from some exceptions ($\pi^+\pi^-,
\pi^0\pi^0, \pi^+\rho^-, \pi^-\rho^+$) our default prediction (with
central values) is again in reasonable agreement with the data. The
agreement is, however, less pronounced than for the pure tree decays
$B^-\to\pi^-\pi^0/\rho^-\rho^0$. Moreover, we point out that the
colour-suppressed modes ($\pi^0\pi^0, \pi^0\rho^0, \rho^0\rho^0$) are
subject to sizeable theoretical uncertainties. This is partly related to
the problem mentioned at the end of the previous section ($\lambda_B$)
and in addition to the fact that these modes are more likely to be
affected by $1/m_b$-corrections.  

\begin{table*}[bt!]
\begin{tabular}{|c|c|cccc|cccc|c|} \hline
\hspace*{2.95cm}&\hspace*{2.3cm}
&\hspace*{1.1cm}&\hspace*{1.1cm}&\hspace*{1.1cm}&\hspace*{1.1cm}
&\hspace*{1.1cm}&\hspace*{1.1cm}&\hspace*{1.1cm}&\hspace*{1.1cm}
&\hspace*{2.3cm} \\[-1.4em]
Mode & Theory & CKM & had & $\mu$ & pow & A & B & C & D & Experiment \\ 
\hline
$B^-\to\pi^-\pi^0$ &
$6.22^{+2.37}_{-2.01}$ &
${}^{+1.14}_{-1.05}$&
${}^{+2.03}_{-1.65}$&
${}^{+0.16}_{-0.18}$&
${}^{+0.43}_{-0.42}$&
$5.97$&
$5.46$&
$6.22$&
$5.64$&
$5.59^{+0.41}_{-0.40}$
\\[0.1em]
$B^-\to\rho_L^-\rho_L^0$ &
$21.0^{+8.5}_{-7.3}$ &
${}^{+3.9}_{-3.5}$&
${}^{+7.4}_{-6.1}$&
${}^{+0.5}_{-0.7}$&
${}^{+1.5}_{-1.4}$&
$20.2$&
$21.3$&
$21.0$&
$23.1$&
$22.5^{+1.9}_{-1.9}$
\\[0.1em]
\hline
$B^-\to\pi^-\rho^0$ &
$9.34^{+4.00}_{-3.23}$ &
${}^{+2.00}_{-1.81}$&
${}^{+3.22}_{-2.51}$&
${}^{+0.31}_{-0.34}$&
${}^{+1.24}_{-0.84}$&
$11.2$&
$10.4$&
$10.3$&
$11.8$&
$8.3^{+1.2}_{-1.3}$
\\[0.1em]
$B^-\to\pi^0\rho^-$ &
$15.1^{+5.7}_{-5.0}$ &
${}^{+2.9}_{-2.8}$&
${}^{+4.8}_{-4.1}$&
${}^{+0.3}_{-0.4}$&
${}^{+1.0}_{-0.7}$&
$11.9$&
$11.9$&
$15.8$&
$11.8$&
$10.9^{+1.4}_{-1.5}$
\\[0.1em]
$\bar{B}^0\to\pi^+\pi^-$ &
$8.96^{+3.78}_{-3.32}$ &
${}^{+1.87}_{-1.91}$&
${}^{+3.02}_{-2.62}$&
${}^{+0.16}_{-0.20}$&
${}^{+1.28}_{-0.71}$&
$6.20$&
$5.21$&
$10.2$&
$5.53$&
$5.16^{+0.22}_{-0.22}$
\\[0.1em]
$\bar{B}^0\to\pi^0\pi^0$ &
$0.35^{+0.37}_{-0.21}$ &
${}^{+0.16}_{-0.14}$&
${}^{+0.20}_{-0.09}$&
${}^{+0.03}_{-0.03}$&
${}^{+0.26}_{-0.11}$&
$0.66$&
$0.63$&
$0.59$&
$0.68$&
$1.55^{+0.19}_{-0.19}$
\\[0.1em]
$\bar{B}^0\to\pi^+\rho^-$ &
$22.8^{+9.1}_{-8.0}$ &
${}^{+4.2}_{-4.0}$&
${}^{+7.8}_{-6.8}$&
${}^{+0.4}_{-0.5}$&
${}^{+1.9}_{-1.4}$&
$20.0$&
$13.2$&
$24.6$&
$15.7$&
$15.7^{+1.8}_{-1.8}$
\\[0.1em]
$\bar{B}^0\to\pi^-\rho^+$ &
$11.5^{+5.1}_{-4.3}$ &
${}^{+2.3}_{-2.1}$&
${}^{+4.2}_{-3.6}$&
${}^{+0.2}_{-0.2}$&
${}^{+1.8}_{-1.0}$&
$13.0$&
$8.41$&
$13.3$&
$11.7$&
$7.3^{+1.2}_{-1.2}$
\\[0.1em]
$\bar{B}^0\to\pi^\pm\rho^\mp$ &
$34.3^{+11.5}_{-10.0}$ &
${}^{+6.3}_{-5.7}$&
${}^{+8.9}_{-7.8}$&
${}^{+0.6}_{-0.7}$&
${}^{+3.7}_{-2.4}$&
$33.1$&
$21.6$&
$37.9$&
$27.3$&
$23.0^{+2.3}_{-2.3}$
\\[0.1em]
$\bar{B}^0\to\pi^0\rho^0$ &
$0.52^{+0.76}_{-0.42}$ &
${}^{+0.10}_{-0.09}$&
${}^{+0.62}_{-0.21}$&
${}^{+0.10}_{-0.10}$&
${}^{+0.41}_{-0.34}$&
$0.44$&
$1.64$&
$0.34$&
$1.02$&
$2.0^{+0.5}_{-0.5}$
\\[0.1em]
$\bar{B}^0\to\rho_L^+\rho_L^-$ &
$30.3^{+12.9}_{-11.2}$ &
${}^{+5.6}_{-5.3}$&
${}^{+11.2}_{-9.6}$&
${}^{+0.6}_{-0.7}$&
${}^{+2.9}_{-2.3}$&
$26.8$&
$22.3$&
$33.2$&
$27.2$&
$23.6^{+3.2}_{-3.2}$
\\[0.1em]
$\bar{B}^0\to\rho_L^0\rho_L^0$ &
$0.44^{+0.66}_{-0.37}$ &
${}^{+0.10}_{-0.09}$&
${}^{+0.50}_{-0.18}$&
${}^{+0.10}_{-0.09}$&
${}^{+0.40}_{-0.30}$&
$0.58$&
$1.33$&
$0.24$&
$1.03$&
$0.69^{+0.30}_{-0.30}$
\\[0.1em]
\hline
\end{tabular}
\parbox{17.9cm}{\caption{\label{tab:BRs}
CP-averaged branching ratios (in units of $10^{-6}$). Experimental
values for $B\to\pi\pi/\pi\rho$ are taken from~\cite{HFAG}, whereas the
ones for $B\to\rho_L\rho_L$ have been inferred from~\cite{rhoLrhoL}. The
different scenarios correspond to: large $\gamma$ (A), large
colour-suppressed amplitude (B), large weak annihilation (C) and a
combined scenario (D). Further details are given in the text.}} 
\end{table*}

In order to illustrate the correlation of the theoretical uncertainties,
we show in Table~\ref{tab:BRs} the central values of some extreme
scenarios (in the spirit of~\cite{Beneke:2003zv}): 

In Scenario~A we study the dependence on the weak phase $\gamma$ (we set
$\gamma=110^\circ$). Modes that show a strong dependence on this
scenario ($\pi^+\pi^-, \pi^0\pi^0, \rho^0\rho^0$) are not particularly
suited for our purposes, as we focus on testing the QCD dynamics of the
topological tree amplitudes in this work. 

In Scenario~B we pursue the question if the data are in accordance with
a large colour-suppressed amplitude, which may be realized in the
factorization framework by a very low value of $\lambda_B=200~\mev$ (we
moreover decrease the form factors to $F_+^{B\pi}(0)=0.21$ and
$A_0^{B\rho}(0)=0.27$). This scenario shows a satisfactory description
of the data, in particular the ''problematic'' modes $\pi^+\pi^-$,
$\pi^{+}\rho^{-}$  and $\pi^{-}\rho^{+}$ are - by construction - in much
better agreement with the data.

It is tempting to understand the large experimentally observed
$\pi^0\pi^0$ branching ratio as an indication for sizeable
non-factorizable power corrections. It is hard to address this issue in
a model-independent way. We would like to emphasize, however, that some
observables are indeed more likely to be affected by $1/m_b$-corrections
than others (cf.~the column labelled ''pow'' in Table~\ref{tab:BRs}). We
in particular expect the branching ratios of the tree decays
$B^-\to\pi^-\pi^0/\rho^-\rho^0$ to be clean observables as they are free
of weak annihilation contributions.

In order to quantify this question we study the influence of a large
annihilation amplitude in Scenario~C (within the BBNS model). It turns
out that it is almost impossible to enhance the $\pi^0\pi^0$ decay rate
and to simultaneously decrease the $\pi^+\pi^-$ rate without fine-tuning
the model parameters~\footnote{It should be noticed that these decay
  rates depend on the same combination of annihilation
  amplitudes.}. Moreover, the overall pattern of branching ratios and in
particular the rates of the other colour-suppressed modes seem to
disfavour a generic scenario with large annihilation contributions. This
is illustrated in Scenario~C, where we double the default value of the
BBNS model for universal weak annihilation, i.e. we set $\rho_A=1$ and
$\phi_A=0$. We conclude that we do not see any clear pattern of
abnormally large power corrections in the data and prefer to be guided
by clean observables rather than by the colour-suppressed and
penguin-contaminated $\pi^0\pi^0$ branching ratio, which cannot be
predicted precisely in the factorization framework anyway. We admit that
our conclusion is a model-dependent statement, which is, however,
supported by a light-cone sum rule analysis, which finds even smaller
annihilation contributions than the BBNS model with default
parameters~\cite{Khodjamirian:2005wn}. 

Finally, Scenario~D is motivated by our analysis of the following
section. It combines elements from Scenario~A and B, but is based on a
more moderate parameter choice $\gamma=90^\circ$, $\lambda_B=250~\mev$
and $F_+^{B\pi}(0)=0.23$, which are within the ranges of our default
parameters from Table~\ref{tab:input}. These values are inspired by a
fit to a set of particularly clean observables that we discuss below. We
refrain from presenting the details of our fit and prefer to simply
illustrate the effects of such a combined scenario~\footnote{We think
  that a sophisticated fit to the observables from
  Table~\ref{tab:ratios} should only be considered when the
  semi\-leptonic $B\to\rho\ell\nu$ spectrum has been measured
  precisely.}.

\section{Precision observables}
\label{sec:test}

Our predictions for the branching ratios from Table~\ref{tab:BRs}
typically have $\sim40\%$ uncertainties, which are largely related to an
overall normalization from $|V_{ub}|F_+^{B\pi}(0)$ and
$|V_{ub}|A_0^{B\rho}(0)$. This particular source of uncertainties can be
eliminated by normalizing the hadronic decay rates to the differential
semi\-leptonic rates at maximum recoil,  
\begin{align}
\frac{d\Gamma}{dq^2}\bigg|_{q^2=0}\hspace{-6mm}
(\bar{B}^0\to\pi^+\ell^-\bar{\nu}_l)
= \frac{G_F^2(m_B^2-m_\pi^2)^3}{192\pi^3m_B^3}
|V_{ub}|^2 |F_+^{B\pi}(0)|^2
\end{align}
and similarly for $\bar{B}^0\to\rho^+\ell^-\bar{\nu}_l$ with
$F_+^{B\pi}\to A_0^{B\rho}$ and $m_\pi\to m_\rho$. The situation is,
however, different for the colour-suppressed modes ($\pi^0\pi^0,
\pi^0\rho^0, \rho^0\rho^0$), which are rather dominated by the
uncertainties from $\lambda_B$ and power corrections than by form factor
uncertainties and $|V_{ub}|$. We therefore do not consider these modes
in this section. 

The BaBar collaboration has measured the semi\-leptonic $B\to\pi\ell\nu$
decay spectrum to high accuracy~\cite{piellnu}. The data has been
investigated in detail under different types of form factor
parameterizations in~\cite{Ball:2006jz}. This analysis uses the HFAG
average for the absolute branching ratio and finds
$|V_{ub}|F_+^{B\pi}(0) = (9.1\pm0.7)\cdot10^{-4}$, which is to be
compared with our default value $10.3\cdot10^{-4}$ and $8.3\cdot10^{-4}$
from Scenario~B. The experimental value has been adopted in conjunction
with our default value for $|V_{ub}|$ to fix the form factor
$F_+^{B\pi}(0)=0.23$ in Scenario~D. 

The analysis of the differential semi\-leptonic $B\to\rho\ell\nu$ decay
spectrum is more complicated as three different form factors contribute
in this case (which confine to $|V_{ub}|A_0^{B\rho}(0)$ at maximum
recoil). Recent measurements by BaBar, Belle and CLEO provide data in
3-4 $q^2$-bins~\cite{rhoellnu}, which does not yet allow to extrapolate
the decay spectrum in a model-independent way. In a recent analysis the
data has been combined with (quenched) lattice calculations of the form
factors in the high $q^2$ region and light-cone sum rule predictions for
$q^2=0$~\cite{Flynn:2008zr}. This analysis yields
$|V_{ub}|A_0^{B\rho}(0) = (5.5\pm2.6)\cdot10^{-4}$, which illustrates
that the data is still premature. We therefore do not include this
number in our analysis. 

Our predictions for the ratios
\begin{align}
{\cal{R}}_{M_3}(M_1   M_2) \equiv
\frac{\Gamma(\bar{B} \to M_1   M_2)}
{d\Gamma(\bar{B}^0\to M_3^+\ell^-\bar{\nu}_l)/dq^2|_{q^2=0}}
\label{eq:slratios}
\end{align}
are shown in Table~\ref{tab:ratios}. For the $\pi\rho$-modes we chose
the normalization such that the dependence on the form factor
multiplying the colour-allowed amplitude is most strongly
eliminated. From Table~\ref{tab:ratios} it can be seen that the
theoretical uncertainties have been reduced considerably to the level of
$\sim15\%$. Moreover, correlations among different sources of
theoretical uncertainties have been resolved to a large extent. 

\begin{table*}[bt!]
\begin{tabular}{|c|c|cccc|cccc|c|} \hline
\hspace*{2.95cm}&\hspace*{2.3cm}
&\hspace*{1.1cm}&\hspace*{1.1cm}&\hspace*{1.1cm}&\hspace*{1.1cm}
&\hspace*{1.1cm}&\hspace*{1.1cm}&\hspace*{1.1cm}&\hspace*{1.1cm}
&\hspace*{2.3cm} \\[-1.4em]
Observable & Theory & CKM & had & $\mu$ & pow & A & B & C & D &
Experiment \\ 
\hline
%
% semileptonic ratios
%
${\cal{R}}_\pi(\pi^-\pi^0)$ &
$0.70^{+0.12}_{-0.08}$ & % default
${}^{+0.01}_{-0.01}$ & % CKM
${}^{+0.11}_{-0.06}$ & % hadronic
${}^{+0.02}_{-0.02}$ & % scale
${}^{+0.05}_{-0.05}$ & % power
$0.68$ & % A
$0.95$ & % B
$0.70$ & % C
$0.82$ & % D
$0.81^{+0.14}_{-0.14}$   % experiment
\\[0.1em]
${\cal{R}}_\rho(\rho_L^-\rho_L^0)$ &
$1.91^{+0.32}_{-0.23}$ & % default
${}^{+0.03}_{-0.04}$ & % CKM
${}^{+0.28}_{-0.17}$ & % hadronic
${}^{+0.05}_{-0.07}$ & % scale
${}^{+0.13}_{-0.13}$ & % power
$1.83$ & % A
$2.38$ & % B
$1.91$ & % C
$2.09$ & % D
$\text{n.a.}$  % experiment
\\[0.1em]
\hline
${\cal{R}}_\rho(\pi^-\rho^0)$ &
$0.85^{+0.22}_{-0.14}$ & % default
${}^{+0.08}_{-0.07}$ & % CKM
${}^{+0.17}_{-0.09}$ & % hadronic
${}^{+0.03}_{-0.03}$ & % scale
${}^{+0.11}_{-0.08}$ & % power
$1.01$ & % A
$1.16$ & % B
$0.93$ & % C
$1.07$ & % D
$\text{n.a.}$ % experiment
\\[0.1em]
${\cal{R}}_\pi(\pi^0\rho^-)$ &
$1.71^{+0.27}_{-0.24}$ & % default
${}^{+0.16}_{-0.18}$ & % CKM
${}^{+0.18}_{-0.12}$ & % hadronic
${}^{+0.03}_{-0.05}$ & % scale
${}^{+0.11}_{-0.08}$ & % power
$1.35$ & % A
$2.07$ & % B
$1.79$ & % C
$1.71$ & % D
$1.57^{+0.32}_{-0.32}$   % experiment
\\[0.1em]
${\cal{R}}_\pi(\pi^+\pi^-)$ &
$1.09^{+0.22}_{-0.20}$ & % default
${}^{+0.15}_{-0.17}$ & % CKM
${}^{+0.03}_{-0.06}$ & % hadronic
${}^{+0.02}_{-0.02}$ & % scale
${}^{+0.16}_{-0.09}$ & % power
$0.75$ & % A
$0.97$ & % B
$1.24$ & % C
$0.86$ & % D
$0.80^{+0.13}_{-0.13}$   % experiment
\\[0.1em]
${\cal{R}}_\pi(\pi^+\rho^-)$ &
$2.77^{+0.32}_{-0.31}$ & % default
${}^{+0.15}_{-0.17}$ & % CKM
${}^{+0.15}_{-0.19}$ & % hadronic
${}^{+0.05}_{-0.06}$ & % scale
${}^{+0.23}_{-0.17}$ & % power
$2.44$ & % A
$2.46$ & % B
$2.99$ & % C
$2.44$ & % D
$2.43^{+0.47}_{-0.47}$ % experiment
\\[0.1em]
${\cal{R}}_\rho(\pi^-\rho^+)$ &
$1.12^{+0.20}_{-0.14}$ & % default
${}^{+0.07}_{-0.07}$ & % CKM
${}^{+0.03}_{-0.06}$ & % hadronic
${}^{+0.02}_{-0.02}$ & % scale
${}^{+0.18}_{-0.10}$ & % power
$1.27$ & % A
$1.01$ & % B
$1.29$ & % C
$1.13$ & % D
$\text{n.a.}$ % experiment
\\[0.1em]
${\cal{R}}_\rho(\rho_L^+\rho_L^-)$ &
$2.95^{+0.37}_{-0.35}$ & % default
${}^{+0.15}_{-0.17}$ & % CKM
${}^{+0.16}_{-0.21}$ & % hadronic
${}^{+0.06}_{-0.07}$ & % scale
${}^{+0.28}_{-0.22}$ & % power
$2.61$ & % A
$2.68$ & % B
$3.22$ & % C
$2.64$ & % D
$\text{n.a.}$ % experiment
\\[0.1em]
%
% hadronic ratios
%
$R(\rho_L^- \rho_L^0/\rho_L^+ \rho_L^-)$ &
$0.65^{+0.16}_{-0.11}$ & % default
${}^{+0.03}_{-0.02}$ & % CKM
${}^{+0.13}_{-0.07}$ & % hadronic
${}^{+0.03}_{-0.03}$ & % scale
${}^{+0.08}_{-0.08}$ & % power
$0.70$ & % A
$0.89$ & % B
$0.59$ & % C
$0.79$ & % D
$0.89^{+0.14}_{-0.14}$ % experiment
\\[0.1em]
$R(\rho_L^+ \rho_L^-/\pi^- \rho^+)$ &
$2.64^{+0.34}_{-0.36}$ & % default
${}^{+0.31}_{-0.31}$ & % CKM
${}^{+0.13}_{-0.13}$ & % hadronic
${}^{+0.00}_{-0.00}$ & % scale
${}^{+0.06}_{-0.14}$ & % power
$2.06$ & % A
$2.65$ & % B
$2.49$ & % C
$2.33$ & % D
$3.23^{+0.69}_{-0.69}$ % experiment
\\[0.1em]
$R(\pi^+ \pi^-/\pi^+\rho^-)$ &
$0.39^{+0.04}_{-0.05}$ & % default
${}^{+0.03}_{-0.04}$ & % CKM
${}^{+0.02}_{-0.02}$ & % hadronic
${}^{+0.00}_{-0.00}$ & % scale
${}^{+0.02}_{-0.00}$ & % power
$0.31$ & % A
$0.39$ & % B
$0.42$ & % C
$0.35$ & % D
$0.33^{+0.04}_{-0.04}$ % experiment
\\[0.1em]
$R(\pi^- \pi^0/\pi^+\pi^-)$ &
$0.65^{+0.19}_{-0.14}$ & % default
${}^{+0.10}_{-0.07}$ & % CKM
${}^{+0.14}_{-0.07}$ & % hadronic
${}^{+0.03}_{-0.03}$ & % scale
${}^{+0.08}_{-0.10}$ & % power
$0.90$ & % A
$0.98$ & % B
$0.57$ & % C
$0.95$ & % D
$1.01^{+0.09}_{-0.09}$ % experiment
\\[0.1em]
$R(\pi^+ \pi^-/\pi^0\pi^0)$ &
$25.7^{+26.0}_{-18.7}$ & % default
${}^{+22.7}_{-10.8}$ & % CKM
${}^{+7.0}_{-11.0}$ & % hadronic
${}^{+2.6}_{-2.3}$ & % scale
${}^{+10.2}_{-10.4}$ & % power
$9.33$ & % A
$8.32$ & % B
$17.3$ & % C
$8.13$ & % D
$3.33^{+0.43}_{-0.43}$ % experiment
\\[0.1em]
\hline
\end{tabular}
\parbox{17.9cm}{\caption{\label{tab:ratios}
Ratios ${\cal{R}}_{M_3}(M_1   M_2) $ of hadronic and differential
semileptonic decay rates as defined in (\ref{eq:slratios}) (in units of
$\gev^2$) and ratios $R(M_1M_2/M_3M_4)$ of hadronic decay rates
from~(\ref{eq:hadratios}). The different scenarios A-D are described in
the caption of Table~\ref{tab:BRs}.}}
\end{table*}

The first two ratios in Table~\ref{tab:ratios} provide particular clean
probes of the QCD dynamics of the topological tree
amplitudes~\cite{facttest,Beneke:2003zv}. In the factorization framework
we have
\begin{align}
{\cal{R}}_{\pi}(\pi^-  \pi^0) \simeq
3\pi^2 f_\pi^2 |V_{ud}|^2 |\alpha_1+\alpha_2|^2,
\end{align}
where small electroweak penguin amplitudes have been suppressed. Our
NNLO prediction for this ratio 
\begin{align}
{\cal{R}}_{\pi}(\pi^-  \pi^0) =
(0.70^{+0.12}_{-0.08})\gev^2
\end{align}
is in good agreement with experimental data 
\begin{align}
{\cal{R}}_{\pi}(\pi^-  \pi^0) |_\text{exp} =
(0.81^{+0.14}_{-0.14})\gev^2,
\end{align}
which strongly supports the factorization assumption. It is, however,
interesting that the central experimental value is in between our
default prediction and the value $0.95~\gev^2$ from Scenario B, which
may hint at a somewhat larger value of the colour-suppressed amplitude
and hence a lower value of the parameter $\lambda_B\simeq~250\mev$
(which we adopt in Scenario~D). Experimental data for the ratio
${\cal{R}}_{\rho}(\rho_L^-  \rho_L^0)$ could help to clarify this
point. 

We recall that all other ratios from Table~\ref{tab:ratios} receive
contributions from QCD penguin amplitudes that are not yet completely
available to NNLO. Among these ${\cal{R}}_{\pi}(\pi^+ \rho^-)$,
${\cal{R}}_{\rho}(\rho_L^+ \rho_L^-)$ and ${\cal{R}}_{\rho}(\pi^-
\rho^+)$ are particularly suited to test the dynamics of the
colour-allowed amplitudes. Our prediction for ${\cal{R}}_{\pi}(\pi^+
\rho^-)$ compares again well to the experimental value. 

The fourth colour-allowed ratio ${\cal{R}}_{\pi}(\pi^+ \pi^-)$ is
special, since the interference of the colour-allowed amplitude with the
QCD penguin amplitude is not negligible in this case. This ratio is thus
particularly sensitive to the choice of the weak phase $\gamma$. One
should keep in mind, however, that the power corrections from weak
annihilation represent another important source of uncertainties for
this ratio. It is interesting to replace the BBNS model for weak
annihilation by the light-cone sum rule prediction
from~\cite{Khodjamirian:2005wn}, which strongly reduces the
uncertainties from weak annihilation and hence enhances the sensitivity
to $\gamma$ (we then find
$1.03^{+0.14}_{-0.16}{}^{+0.03}_{-0.06}{}^{+0.02}_{-0.02}{}^{+0.03}_{-0.03}~\text{GeV}^2$).
The current experimental value may then be considered as a hint at a
large value $\gamma\gtrsim90^\circ$. A smaller value of $\gamma$, on the
other hand, may then imply the presence of an additional contribution to
the QCD penguin amplitude or that power corrections, which are neither
from weak annihilation nor from chirally enhanced wave functions, have
been underestimated in our approach. The latter would be conceptually
important, as it would increase the total uncertainty from power
corrections in QCDF. We refrain, however, from drawing any conclusions
concerning $\mathcal{R}_\pi(\pi^+\pi^-)$ and its implications for
$\gamma$, as long as the penguin amplitudes have not been calculated to
NNLO. 

In Table~\ref{tab:ratios} we also show some ratios of hadronic decay
rates defined by 
\begin{align}
R(M_1 M_2/M_3 M_4) \equiv
\frac{\Gamma(\bar{B} \to M_1 M_2)}
{\Gamma(\bar{B}' \to M_3 M_4)}.
\label{eq:hadratios}
\end{align}
The ratio $R(\rho_L^- \rho_L^0/\rho_L^+ \rho_L^-)$ yields complementary
information on the tree amplitudes from the $\rho$-sector, where the
contamination from the QCD penguin amplitudes is known to be less
important~\cite{Aleksan:1995wn,QCDF:VV}. We consider the experimental
value for this ratio as another important evidence in favour of an
enhanced colour-suppressed amplitude (Scenario B or D).

The ratios $R(\rho_L^+ \rho_L^-/\pi^- \rho^+)$ and $R(\pi^+
\pi^-/\pi^+\rho^-)$ of colour-allowed modes can be predicted precisely
in the factorization framework. Whereas the second ratio is in nice
agreement with the data, the first one seems to somewhat disfavour a
scenario with a large weak phase $\gamma$. 

The last two ratios from Table~\ref{tab:ratios} finally refer to what is
known as the $B\to\pi\pi$ puzzle. Whereas the ratio $R(\pi^-
\pi^0/\pi^+\pi^-)$ is by construction in Scenarios~A, B and D in better
agreement with experimental data than our default prediction, the ratio
$R(\pi^+ \pi^-/\pi^0\pi^0)$ illustrates what we mentioned at the
beginning of this section, i.e.~the bulk of theoretical uncertainties
does \emph{not} drop out in ratios that involve colour-suppressed
modes. The uncertainties of our default prediction
\begin{align}
R(\pi^+ \pi^-/\pi^0 \pi^0)
= 25.7^{+26.0}_{-18.7}
\end{align}
are thus extremely large and the central value is in vast disagreement
with the data. This ratio may be brought down by a factor of $\sim3$ in
Scenarios~A, B and D, which may be seen as an independent evidence in
favour of these scenarios. The fact, however, that these predictions
still suffer from $\sim60\%$ uncertainties related mainly to the power
corrections, $a_2^\pi$, $\mu_{hc}$ and $f_B$ (in decreasing order of
importance), shows that we cannot expect to predict this ratio
precisely. We would like to add that there is no such puzzle in the
$\pi\rho/\rho\rho$ channels, i.e.~there is no general failure of QCDF to
describe colour-suppressed modes.

\section{Conclusions}

We presented the NNLO QCDF prediction for the approximate tree decays
$B^-\to\pi^-\pi^0/\rho^-\rho^0$ and updated the global analysis of the
other tree-dominated $B\to\pi\pi/\pi\rho/\rho\rho$ decay modes. Our
analysis from Section~\ref{sec:test} showed that QCDF yields precise
theoretical predictions for particular \emph{ratios} of decay rates. We
find in general support for the factorization assumption and uncovered
some hints for enhanced colour-suppressed amplitudes, which translate in
QCDF into a small value of the $B$ meson parameter
$\lambda_B$. Theoretical progress from non-perturbative methods on the
hadronic ratio $f_{M_1}\hat{f}_B/\lambda_BF^{B M_1}(0)$ as well as
experimental measurements of the semileptonic $B\to\rho\ell\nu$ decay
spectrum may shed further light on this issue.

\begin{acknowledgments}
We are grateful to Gerhard Buchalla for interesting discussions and
comments on the manuscript. We would like to thank Andreas H\"ocker for
helpful correspondence. The work from G.B.~was supported by the DFG
Sonderforschungsbereich/Transregio~9. The work from V.P.~is partially
supported by the Swiss National Foundation as well as EC-Contract
MRTNCT-2006-035482 (FLAVIAnet). The Albert Einstein Center for
Fundamental Physics is supported by the ''Innovations- und
Kooperationsprojekt C-13 of the Schweizerische Universit\"atskonferenz
SUK/CRUS''.  
\end{acknowledgments}

\end{document}